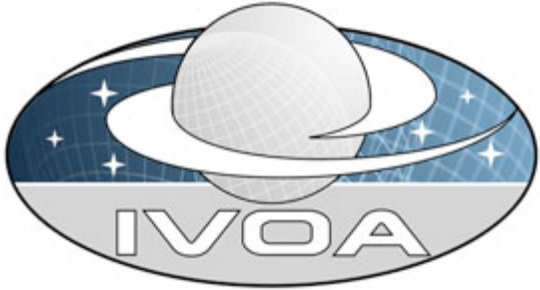

**International**

**V**irtual

**O**bservatory

**A**lliance

AN ENCODING SYSTEM TO REPRESENT STELLAR SPECTRAL CLASSES IN ARCHIVAL DATABASES AND CATALOGS

# 1 Version 1.04

*Design Note: 2011 December 15*

**This version:**
    1.04-20111215

**Previous version(s):**
    1.03-20090312
    1.02-20081027
    1.01-20080904

**Change log from V1.03**

- Change of definition of tt=20 for OBA stars (Table 1)
- Revision of Notes to Table 3, white dwarfs P3 & P4
- Treatment of composite strings: "+...", ".","comp."
- Clarification of meanings of "/" symbol.
- Updating of Figure 1.
- Define "global" classes as nonexploding stars for peculiarities.

**Editor(s):**
    Myron A. Smith

**Author(s):**
    Myron A. Smith, Randall W. Thompson, Richard O. Gray, Chris Corbally, Inga Kamp

### *Abstract*

The data archives from space and ground-based telescopes present a vast opportunity for the astronomical community. We describe a classification encoding system for stellar spectra designed for archival databases that organizes the spectral data by "spectral classes." These classes are encoded into a digital format of the form **TT.tt.LL.PPPP**, where **TT** and **tt** refer to spectral type and subtype, **LL** to luminosity class, and **PPPP** to possible spectral peculiarities. Archive centers may wish to utilize this system to quantify classes of formerly arbitrary spectral classification strings found in classification catalogs corresponding to datasets of pointed spectroscopic observations in their holdings. The encoding system will also allow users to request archived data based on spectral class ranges, thereby streamlining an otherwise tedious data discovery process. Material in Appendix A is "normative" (part of the defined standard). Appendices B and C are "informative," meant to show how one data provider (MAST) has opted to handle some practical details.

### *Status of This Document*

An earlier version of this Working Draft was posted on the IVOA web site at http://www.ivoa.net/Documents/, where a list of *current IVOA Recommendations and other technical documents* can also be found. This document was discussed for promotion at the IVOA Interoperability meeting in Baltimore in late October of 2008. The current version of the document has been reorganized to develop further the peculiarity codes for less commonly used spectral types and to address newly discovered details for this standard as well as practical matters of implementation in Appendices B and C.

*This Note is intended for discussion among the cognizant IVOA committees and has been presented formally to the IVOA semantics Committee for promotion at the October, 2008 IVOA Interoperabilty meeting in Baltimore. It is also intended for other data providers that have discovered the need for a spectral class nomenclature system in their own databases for spectroscopic missions. It has also been read and vetted by designated astronomers who are experts in the field of spectral classification and who have helped clarify how to best translate spectral types into a format useful for retrieval of information from stellar databases. The parsing code written to implement the nomenclature system, written in PHP, is available to other data providers upon request.*

### *Acknowledgements*

We thank Dr. Robert Hanisch for his advice in the writing and dissemination of this document to the cognizant IVOA groups, to Dr. Rick White of MultiMission Archive at Space Telescope (MAST) for his encouragement, and members of the IAU Commission 45 (Stellar Classification) for their review of this document.

## 2 Contents

## 3  Introduction

Astronomers using archival data have a need to download metadata (descriptive parameters of objects) or data (FITS files describing wavelengths and fluxes) for groups of like stars without making explicit searches by star name one at a time or by cross correlation of established catalogs. This is a common request to archive centers holding spectroscopic datasets. NASA's International Ultraviolet Explororer (IUE) satellite project, with the Far Ultraviolet Spectroscopic Explorer (FUSE) project following suit some years later, attempted to solve this problem by cataloging stars by "Object Classes."  Later NASA space-borne missions have not attempted any such plan and at most have provided "object classes" as arbitrary strings taken from the literature. Then for all practical purposes each star's entry becomes a group containing as few as one member.

There were two general problems that prevented the IUE Object Class system from succeeding and thereby preventing archival data users from conveniently retrieving archival information on groups of stars with common properties. One was that, owing to its simplified nature, the object class scheme introduced ambiguities that limited its usefulness. For example, a B2Ve star could be called either a B2V star (thus, B0-2 II-V spectral types correspond to Object ID = 20) or a Be star (Class 26). Since both descriptions are accurate but incomplete, it often happened (in this example) that different observations of the same object catalogued this ID number as either 20 or 26. Therefore, searching on all the observations of Class 20 would not give a complete list of the observation identifiers.  More generally, the scheme was limiting in that it did not require a single consolidated resource of spectral types.

The need to develop a classification system for retrieving astrophysically relevant data for celestial objects has been anticipated by the IVOA. For example, the Spectrum Data Model (Version 1.1; see Section 5.6) states:

*"At the moment there is no international standard list of valid values for Target class and spectral class. Nevertheless an initial deployment of the VO would gain some benefit from using archive-specific classes, and provide a framework for converging on a standard list."*

This document addresses this need by describing a new nomenclature system of "spectral classes" for stars. This system must be broad enough to account for well defined spectral peculiarities and yet must be conceptually easy enough for readers to understand from a brief description, if indeed they need to understand it at all. A byproduct of this nomenclature system is that it could provide a conceptual framework for a more universal system that includes types of astronomical objects other than stars. Our scheme is defined in Section 4 and is developed by examples of single star and composite spectra in Section 5 to demonstrate both conventional and unconventional classification details that astronomers have constructed over the years. Appendix A goes into necessary detail about nonstandard classification string structures for certain peculiar stars like Am stars. Appendices B gives examples of web interface forms that an archive center might develop to exploit this system while Appendix C discusses optional information in the form of a "detail" flag designed to give supplementary information.

# 4 The Concept of "Spectral Classes"

We define the term "spectral class" as a cell in three dimensional parameter space that describes a star's (usually optical) spectrum in terms of its spectral type, luminosity class, and possible well defined peculiarities. A star's spectrum may be uniquely and unambiguously assigned to a spectral class cell by means of an ASCII string known as its spectral classification. A spectral class system is a series of digital codes that expresses this classification. As mentioned in the previous paragraph, the concept of a spectral class for astronomical objects was anticipated in the IVOA Spectral Data Model document. To take this concept a step further, we describe the mapping of spectral classes in the IVOA context by means of a series of integer codes by which spectral types of **stars** can be substantially specified by the above three classification parameters. A key purpose of such a system is to facilitate large and targeted retrievals of information from any astronomical data center.

## *4.1 General Guidelines*

The general philosophy behind a spectral class nomenclature scheme is that it adheres to the following guidelines:

i. All attributes are entirely derivable from a star's spectrum (generally in the blue-optical region, but if need be from the near-IR or near-UV), and not from the star's photometric characteristics (e.g., color or variability).
ii. The overall spectral class system should adhere to general rules and conventions of stellar spectral classification for like groups of stars.
In contrast, they should not be taken from an *ad hoc* spectral type contrived for a particular star's peculiar and perhaps unique spectrum. We will develop this point by listing in Table 1 the set of commonly defined spectral types.
iii. A star's spectral class should be uniquely mapped to elements of the scheme. (Reverse mapping is also desirable, but it is too difficult to achieve because of the multiple values given in the table fields in Tables 1-3.)
iv. The coding system should provide room for potential growth and development for possible discovery of new spectral types or peculiarities.
v. The scheme should ignore nonstandard spectral type or peculiarity descriptions. (See Section 4.6.)

Certain guidelines, (e.g. point ii) will be left deliberately vague because spectral types are human contrivances. They are best left to the consensus of experts who are aware of the needs of classification criteria to be static on one hand and of infrequent additions of attributes on the other. We have recruited as experts the Organizing Committee of the IAU Commission 45 (Stellar Classification) and two authors of this Note, Richard Gray and Christopher Corbally, who have recently written a monograph on spectral classification (see References). We anticipate that when canvassing new spectral classification catalogs, classification strings for the certain classes of stars will have evolved. This will require occasional changes in our standards and table definitions.

## *4.2 Scheme outline:* TT.tt.LL.PPPP

We specify the format of our spectral class system by the string **TT.tt.LL.PPPP**. Each letter is represented as an integer from 0 through 9. In a database table. However, experience has shown that storing each of the seven values (i.e., **TT, tt, LL, $P_1$, $P_2$, $P_3$, $P_4$**) as separate integer fields allows more flexible query options (e.g., range searches and other mathematical operations).

## 4.3 Representation of spectral types and luminosity classes

Pairs of digits for **TT**, **tt,** and **LL** may range from 00 through 99. These 100 numbers are cells denoting spectral types, subtypes, and luminosity classes respectively, as in the Morgan-Keenan (MK) system, - and beyond. For all four major classification parameters, including spectral peculiarities, a "00" or "0" represents "no information." In order to allow for both traditional major spectral types and some new ones, a second digit is allocated. In addition, the allotment of two digits for **LL** permits subdivision of more than the traditional I-V luminosity subclasses, especially for supergiants and stars near the Zero Age Main Sequence. Tables that map types and class to the **TT.tt.LL** digits are given in Tables 1 and 2. We note parenthetically that apart from spectral types L and T, such types as the hypothetical "Y" (theoretical, with $T_{eff}$ < 700°K) are offered to provide possible growth for "undiscovered" spectral types. The TT.tt values are understood to refer to hydrogenic types, which in a few cases may be designated with the character "h" preceding the spectral type (see Section 4.6).

We note in particular that decimal subtypes are encoded only for subtypes 9.5 of O and B stars (including DO, DB, WO, WB) in the standard. Users interested in decimal subtypes may consult Appendix C as one way in which this information can be flagged.

# Table 1. Mapping of TT spectral types and tt subtypes

ENCODED SPECTRAL TYPES (TT):

| Sp | Code | Description |
|---|---|---|
| -- | 00 | Unknown |
| O, OC, ON | 10 | O type |
| B, BC, BN | 11 | B type |
| A | 12 | A type |
| F, sdF | 13 | F type |
| G, sdG | 14 | G type |
| K, sdK | 15 | K type |
| M | 16 | M type |
| L | 17 | L type |
| T | 18 | T type |

| Sp | Code | Description |
|---|---|---|
| Y | 19 | For future cool brown dwarfs |
| C | 20 | Undifferentiated C stars |
| CR | 21 | R carbon stars |
| CN | 22 | N carbon stars |
| CJ | 23 | J carbon stars |
| CH | 24 | CH stars |
| CHd | 25 | Hydrogen deficient carbon stars |
| S, SC, MS | 26 | S-type stars |
| CV | 27 | Novae |
| ?? | 28 | Open for future |

| Sp | Code | Description |
|---|---|---|
| SN | 29 | Supernovae |
| sd | 30 | subdwarf without type |
| sdO, sdON, sdOC, sdOB | 31 | O-type subdwarfs |
| sdB, sdBN, sdBC | 32 | B-type subdwarfs |
| sdA | 33 | A-type subdwarfs |
| D, WD, wd | 40 | white dwarf without type |
| DA | 41 | white dwarf with hydrogen lines |
| DB | 42 | white dwarf with He I lines |
| DC | 43 | white dwarfs with continuous spectrum |
| DO | 44 | white dwarf with He II lines |
| DZ, DF, DG, DK, DM, DX | 45 | White dwarf with metal lines |

| Sp | Code | Description |
|---|---|---|
| DQ | 46 | white dwarfs with carbon lines |
| PG | 47 | PG1159 stars |
| D? | 48 | reserved for new WD type |
| D?? | 49 | reserved for new WD type |
| NS | 50 | neutron stars |
| WR | 51 | Wolf Rayet without type |
| WN | 52 | N-sequence Wolf Rayet |
| WC | 53 | C-sequence Wolf Rayet |
| WO | 54 | O-sequence Wolf Rayet |
|  | 55 | Open for new Wolf Rayet type |
| OB, OBN, OBC | 56 | OB stars |

**NOTES:**

1. This table omits some obsolete usages. For example, R and N map to CN and CR, respectively. The scheme interprets "R" and "N" as these respective spectral types.

2. For the reader's information (only), subdwarf A stars (sdA → TT=33) consist of two evolutionarily distinct groups: sdA0-sdA2 stars are products of advanced evolution (like sdO, sdB), whereas sdA3..sdA9 stars (like sdF-sdK stars) are old stars on the metal-weak main sequence.

3. "Undifferentiated C stars" are not a universally sanctioned spectral type, but because these designations continue in the literature they have been added to our table; TT=20.

*Boldface are "base" classes, to which plain font classes are considered aliases.
Ignore "–" and "+" in superscripts, as in "Va⁻", "Va⁺". Retain "+" for Ia⁺ only.
'esd' is used as a luminosity class for F, G, and K stars but as a peculiarity for M, L,
and T stars. In both cases it immediately precedes the spectral type.*

ENCODED SUBTYPES (**tt**) FOR NON-EXPLODING STARS:

| Code | Subtype |
|---|---|
| 00 | Unknown |
| 10-19 | 0-9 |
| 20 | 9.5-9.10 for hot stars (TT=10, 11, 31, 32, 52-54, 56) |
|  | 10 for other spectral types (e.g. A-M) |

ENCODED SUBTYPES (**tt**) FOR NOVAE AND SUPERNOVAE:

| Code | Nova Class |
|---|---|
| 00 | Unknown |
| 10 | He/N |
| 11 | Fe IIn |
| 12 | Fe IIb |

| Code | SN Class |
|---|---|
| 00 | Unknown |
| 10 | Ia |
| 11 | Ib |
| 12 | Ic |
| 13 | Ib/Ic |
| 14 | II |
| 15 | IIb |
| 16 | IIn |

## Table 2. Mapping of LL luminosity classes

| Class | Code | Class | Code | Class | Code |
|---|---|---|---|---|---|
| Unknown | 00 | **IIa** | 18 | **IV**, IVab, III-V, III/V | 27 |
| **0** | 10 | **II**, IIab, I-III, I/III | 19 | **IVb** | 28 |
| **0-Ia**, Ia0, Ia-0, Ia⁺ | 11 | **IIb** | 20 | **IV/V**, IV-V, IVa-V, IVa/V | 29 |
| **Ia**, Ia0-Ia, Ia0/Ia | 12 | **II-III**, II/III, IIb-III, IIb-IIIa | 21 | **Va**, V, Va⁺, Va⁻, d, Va-V, Vab | 30 |
| **Ia-Iab**, Ia/Iab, Ia-ab | 13 | **IIIa** | 22 | **Vb**, Vz, Vb-Vz | 31 |
| **Iab**, I | 14 | **III**, IIIa-III, IIIab, III-IIIa, g | 23 | **VI** | 32 |
| **Iab-Ib**, Iab/Ib, Iab-b | 15 | **IIIb**, III-IIIb | 24 | **VII**, esd | 33 |
| **Ib**, I-II, I/II | 16 | **III-IV**, IIIb-IV, IIIb-IVa, III/IV | 25 | **VIII** | 34 |
| **Ib-II**, Ib-IIa, Ib/II | 17 | **IVa** | 26 | **IX** | 35 |

# Table 3. Mapping of $P_1P_2P_3P_4$ spectral peculiarities

Table of "Global" Peculiarities $P_1$ and $P_2$ (except for novae, CV, supernovae)

Code    Pec string    Description

| Code | Pec string | Description |
|---|---|---|
| 0 | | Unspecified |
| 1 | + | Composite (P3,P4 describe secondary) |
| 2 | p, pec | Peculiarity |
| 3 | E, em | Emission lines |
| 4 | [e], q | Forbidden line emission |
| 5 | v, var | Variable |
| 6 | s | Sharp lines |
| 7 | n, nn | Broad lines |
| 8 | | Open for future |
| 9 | | Open for future |

### NOTES ON "GLOBAL" PECULIARITIES (P1P2):

1. If two or more global peculiarities are present in the classification string, only the two with the lowest values in this table are selected.
2. If one global peculiarity is present, its value is assigned to P1 and P2 is given the value 0.
3. The values of two nonzero global peculiarities will be mapped to P1 and P2 in order of their values in this table (i.e. P1 < P2). The exception is that if "p" and "e" both appear they retain their original classification order.
4. For composite spectra the following rules apply:
   a) TT.tt.LL.P1P2 codes refer to the primary's spectrum, and P1=1.
   b) P3P4 refer to the spectral type of the secondary's spectrum.

## Table of Peculiarities P3, P4 determined by TT Spectral types

### Wolf-Rayet Stars (TT = 51-55)

| P3 | Pec | Description |
|---|---|---|
| 0 | | No peculiarity |
| 1 | b | broad-lined |
| 2 | (h) | weak hydrogen |
| 3 | h | hydrogen emis. present |
| 4 | ha | hydrogen emis + abs present |

| P4 | Pec | Description |
|---|---|---|
| 0 | | No peculiarity |

### Types OB (also OB, sdB, sdO, sDA; TT = 10, 11, 31, 32, 33, 56)

| P3 | Pec | Description |
|---|---|---|
| 0 | | No peculiarity |
| 1 | He | He strong or He weak |
| 2 | HgMn, Hg, Mn | Hg and/or Mn |
| 3 | Si, SiSr | Si or Sr strong |
| 4 | SiCrEu, SiCr | SiCrEu or SiCr strong |
| 5 | SrCrEu | SrCrEu strong |
| 6 | SiEu | SiEu strong |
| 7 | Cr | Cr strong |
| 8 | Sr, SrSi | Sr or SrSi strong |

| P4 | Pec | Description |
|---|---|---|
| 0 | | No peculiarity |
| 1 | PCyg, w | wind |
| 2 | f | emission line class |
| 3 | f+ | emission line class |
| 4 | f* | emission line class |
| 5 | Fe+, Fe, m+, m | metal strong |
| 6 | Fe-, m-, wk | metal weak |
| 7 | sh, shell | shell lines |
| 8 | C, N | CN peculiarities are triggered by above TT's |

Note to previous table:
1. For O, B, OB, sdO, sdB stars, C, N peculiarities may immediately follow the spectral type, e.g. "OC9 V".
2. These peculiarities strings and others in the following P3P4 tables may be denoted in different but generally obvious ways by some classifiers, e.g. "m" may be "metallic," etc.
3. The "He" anomaly is recognized for B stars. In rare cases when "He" precedes type A, these characters are ignored.

Types AF (also sdF; TT = 12, 13)

| P3 | Pec | Description |
|---|---|---|
| 0 |  | No peculiarity |
| 1 | HgMn, Mn, Hg | Hg and/or Mn strong |
| 2 | Si | Si strong |
| 3 | SiCrEu | SiCrEu strong |
| 4 | SrCrEu | SrCrEu strong |
| 5 | SiEu | SiEu strong |
| 6 | Cr | Cr strong |
| 7 | Sr | Sr strong |

| P4 | Pec | Description |
|---|---|---|
| 0 |  | No peculiarity |
| 1 | P Cyg | P Cygni profiles |
| 2 | lam Boo, l Boo, lambda Boo | lambda Boo star |
| 3 | Fe-, m-, w, wk | metal-weak or subwarf |
| 4 | Fe+, m+ | metal-strong |
| 5 | m | metallic lined |
| 6 | sh, shell | shell lines |
| 7 | Ba | Ba dwarf |
| 8 | rho Pup | rho Puppis star |

Note:
1. The P4=3 "Fe-" anomaly is triggered automatically for type sdF, as for metal-weak stars.

Types GK (also sdG, sdK; TT = 14, 15)

| P3 | Pec | Description |
|---|---|---|
| 0 |  | No peculiarity |
| 1 | CN+, CN | strong CN |
| 2 | CN- | weak CN |
| 3 | CH+, CH | strong CH |
| 4 | CH- | weak CH |
| 5 | CN+ CH+ | (self-explanatory) |
| 6 | CN+CH- | (self-explanatory) |
| 7 | CN- CH+ | (self-explanatory) |
| 8 | CN- CH- | (self-explanatory) |

| P4 | Pec | Description |
|---|---|---|
| 0 |  | No pecuiarity |
| 1 | Ba+, Ba | barium star |
| 2 | Ba Fe+ | metal-strong barium star |
| 3 | Ba Fe- | metal-weak barium star |
| 4 | Fe+, m+ | metal-strong |
| 5 | Fe-, m- | metal-weak or subdwarf |
| 6 | C2 | strong Swan bands |
| 7 | Ba C2 | Ba star, strong Swan bands |
| 8 | Ca | calcium peculiarity |

Note:
1. The P4=5 "Fe-" anomaly is triggered for the sdG, sdK types, as for metal-weak starrs.

Types MLT (TT = 16-18)

| P3 | Pec | Description |
|---|---|---|
| 0 |  | No peculiarity |
| 1 | Ba+, Ba | Ba strong |
| 2 | sd | subdwarf |
| 3 | esd, usd | Extreme or ultra subdwarf |

| P4 | Pec | Description |
|---|---|---|
| 0 |  | No peculiarity |
| 1 | Fe+ | Fe strong |
| 2 | Fe- | Fe weak |

## White Dwarfs ("D" types TT = 40-49)

| P3, P4 | Peculiarity | Description |
|---|---|---|
| 0 | | No peculiarity |
| 1 | A | trace H I lines |
| 2 | B | trace He I lines |
| 3 | O | trace He II lines |
| 4 | Q | trace carbon lines |
| 5 | Z | trace metal lines |
| 6 | H | magnetic, w/o visible polarization |
| 7 | P | magnetic with visible polarization |
| 8 | d | dust |

Notes:
1. As noted by Gray & Corbally (2009): "To the primary type can be added one or more secondary composition symbols
   (A, B, C, O, Z, or Q), indicating a trace of an element defined as for the primary ones [spectral types]." These compositional symbols as well as the P and d designations are also defined In the table.
2. The same symbol may not be used for both spectral type and compositional purposes, e.g. "DAA" is not used.

## Type C* (Carbon Stars, TT = 20-25)

| P3 | Pec | Description |
|---|---|---|
| 0 | | No peculiarity |
| 1 | MS | Merrill-Sanford bands |
| 2 | dC | Dwarf carbon star |

| P4 | Pec | Description |
|---|---|---|
| 0 | | No peculiarity |
| 1 | j | enhanced C13 |

## Type S (S Stars, TT = 26)

| P3 | Pec | Description |
|---|---|---|
| 0 | No peculiarity | No peculiarity |
| 1 | MS | MS (marginal S-type) stars |
| 2 | /1, /2, /3 | CO index= 1,2 or 3 |
| 3 | /4, /5, /6 | CO index =4, 5 or 6 |
| 4 | SC, /7, /8, /9, /10 | CO index =7,8,9 or 10 |

| P4 | Pec | Description |
|---|---|---|
| 0 | | No peculiarity |
| 1 | Tc+, Tc | Tc-strong |
| 2 | Tc- | Tc-weak |

Notes affecting C and S star peculiarities:
1) There are two possible "peculiarity triggers for TT=26 (S and related) stars:
   a) "SC" is a spectral type that triggers an SC anomaly, namely P3 = 4.
   b) "MS" is a spectral type that triggers P3=1.
2) "MS" in the C and S star tables has two different meanings, as follows:
   *a)* for C stars it refers to the presence of Merrill-Sanford (Si-C-C) bands; typically "MS" occurs at the end of a classification string.
   *b)* for S stars it refers to a marginal S star.

## 4.4 Assignment of Spectral Peculiarities, $P_1P_2P_3P_4$

All stars including the standards are peculiar at some level. Hence, the most complicated part of any classification system, is the array of spectral peculiarities, usually designated by a single letter or character. The descriptions of these attributes, $P_1P_2P_3P_4$, each of which has a value in the range 0-9 as well as Notes for them are given in Table 3. The integers $P_1$ and $P_2$ encode spectral peculiarities that can appear "globally" across large regions of the HR Diagram (but not exploding stars). They are mutually duplicative in order to represent as many as two global peculiarities. Codes $P_3$ and $P_4$ are included to specify more fully the peculiarities that often cannot be represented by $P_1$ and $P_2$ alone. Peculiarities $P_3$ and $P_4$ often refer to chemical abundance anomalies, and they tend to differ from one spectral type to another, and each of their values is distinct from any other. So, for example, a value $P_4$ = 3 for A star spectra has a different meaning than $P_4$ =3 for M stars. Table 3 lists these codes for each major spectral type.

In order to maintain flexibility of user choice, it is necessary to impose a few rules to the $P_1$ and $P_2$ codes at the expense of simplicity. The first of these is that if only one global peculiarity is present, its value is assigned to $P_1$. Second, with one exception, the peculiarities encoded by $P_1$ and $P_2$ are assigned according to the order in which they appear in the $P_1P_2$ table in Table 3. So, for example, "Aen" and "Ane" stars will be given the same $P_1$ and $P_2$ codes ($P_1P_2$ = 37). The exception to the rule of ordering by table value occurs only when the peculiarities e and p both occur in the classification string. The ordering is then "noncommutative," i.e, p and e are placed $P_1$ and $P_2$ in order of their appearance in the classification string.

Notice also that the string "B4Vnvarp" is interpreted as "B4Vpvar": first, because p has a lower value than n or var and must be placed first among the $P_1P_2$ peculiarities and, second, because n has the greatest numerical value of the three, it will be discarded.

The $P_3$ and $P_4$ anomalies are normally used to further clarify what is meant by "peculiar" (see Table 3). However, for composite spectra (identified by $P_1$ = 1) $P_3$ and $P_4$ designate the spectral type of the secondary star – that is, this two-digit code gives the **TT** type of the secondary. If the secondary's type is not known, it is coded in $P_3P_4$ as "00".

We handle the precedence of the $P_3$ and $P_4$ peculiarities in left to right order as they appear in the classification string. If more than two such peculiarities are present only the left-most two values are interpreted.

In a very few cases the spectral type itself will trigger a spectral peculiarity. This follows conventions already adopted by certain classification communities. Examples of this include sdF, sdG, sdK stars which trigger an "Fe-" **P$_4$** peculiarity. However, the intention of our scheme is to reduce to a minimum the instances of "implied peculiarities."

Peculiarity designations, as some spectral types, are case-dependent.

## 4.5 "Peculiar Exceptions"

We note that special cases occur wherein the classification does not follow the convention "spectral type + luminosity class + peculiarity" because of the location in the string where the peculiarity is represented. Departures from this convention may then be required, as discussed fully in Appendix A. For example, some classifiers publish spectral types like "kA5hF0mF2" or even "A5/F0/F2" to represent spectral types/subtypes to Am or Fm stars according to the placement of the Ca K, hydrogen, and "metallic" line symbols. In such cases the k and the h symbols will be ignored and the middle type (pertaining to the hydrogen lines) will be represented in the nomenclature code. In general, the k, h, and m anomalies are interpreted as preceding the subtype in Am stars. (In our two examples, the strings would be represented as "F0m".)

## 4.6 Special Characters

Some types of complex spectral type strings still exist that cannot be easily mapped to this coding system. Because of this fact, the coded classification cannot always be reverse-mapped back to the original string. In order to avoid further complexity, some characters that have well understood meanings will be ignored. Examples of ignored symbols or abbreviations are "?", ":", "abs", "…", "g" (referring to Fraunhofer G-band type), and (generally) "(" and ")". The sole exception to the () symbols occurs for the anomaly "(ha) for Wolf-Rayet stars (see Table 3). Other special characters are interpreted as follows:

- . (decimal point)   A number after a decimal point, intended by the original classifier as a decimal spectral subtype **tt,** is generally ignored and not parsed, e.g. B0.2 is parsed as B0. However, as specified for **tt** = 20 for hot stars in Table 1, the decimal subtype is not ignored for subtypes 9.5-9.10. Thus, B9.7 is parsed as B9.5. For parsing of spectral types from the Simbad catalog a single "." is ignored, e.g. SN.Ia → SN Ia.
- - (dash symbol)   This symbol between two Roman numerals is interpreted as an intermediate luminosity class. See the luminosity class table in Table 2. Note the "-" may also represent a minus, as for certain peculiarity strings noted in Table 3. Also, if "-" occurs between two Arabic numbers, spectral types, or subtypes, the first value is selected.

- / (slash symbol) If this symbol appears between two Roman numerals, as with "Iab/I", it is treated just like a dash symbol. This usage is defined in the LL table of Table 2. The note just above for "-" separating numbers, etc. applies here too. For example, G8/K0 becomes G8 and G8/K0/III/IV becomes G8III. Two exceptions to this convention include the "/ notation used by some classifiers for Am stars (see Appendix A) and the inclusion of a C/O index for S stars (see Table 3).

- + (plus sign) Between two spectral type substrings this character signifies a composite spectral type. For this purpose the string "+…" (SIMBAD syntax for composite spectrum) has a clear meaning and should be parsed to give the peculiarity code "1000." Also, sometimes the "+" can appear as a modifier to a luminosity class (such as "Ia+") or spectral type (such as OB+). In our scheme the "+" will be dropped anywhere but for the composite spectrum designation. Valid peculiarities using a "+" symbol are described in Table 3.

- "h" (lower case) this character is unparsed, except it signals that a spectral type (generally A or F) immediately follows it. (See also Appendix A.)

These symbols or other non-alphanumeric characters will be ignored if they appear in other contexts.

# 5 Examples

We exhibit many of the facets of the coding system by means of examples that bring out different general and sometimes subtle aspects of the system.

## 5.1 Single-star spectra

- F5

We expect that most databases will contain many stars for which detailed information on the spectral type is unavailable. Thus, a spectral type of "F" alone would be encoded as: F → 13.00.00.0000. All zeroes signify an absence of classification information. They do not speak to the possibility that a new classification based on a higher quality spectrum than was used to determine the F5 classification might disclose a luminosity class and peculiarities. It follows that F5 → 13.15.00.0000, where, from Table 1, code 15 represents the subtype 5. The implied ambiguity for luminosity classes and peculiarities (as unknowns) in this example is handled by the 0 values.

- A3 IVn

A3 IVn → 12.13.27.7000. This example represents the **LL** code 27 for luminosity class IV and the peculiarity code $P_1$ =7 for "n" (nebulous; i.e. broad-lined) – see Table 3. Here a nonzero $P_1$ value is assigned because an 'n' peculiarity may be defined for all spectral types. Also note that there is no distinction between "(n)," "n," and "nn."

- O7 IIIf

O7 IIIf → 10.17.23.0002. This example differs from the previous one in that "f" is a spectral type-dependent peculiarity, being defined only for O-type stars. The code value for this peculiarity, $P_4$ = 2, is given in Table 3.

- SN Ia

SN Ia → 29.10.00.0000 . The **TT** table in Table 1 gives a spectral type code for all supernovae, 29. We emphasize the point that the Roman numeral "I" (denoting a type of supernovae) does not represent a luminosity class on the HR Diagram. Rather, it serves to bifurcate the phenomenological types of SNs. Because "Ia" is interpreted as the **tt** spectral subtype, Ia → 10.

- sdO

sdO → 31.00.00.0000. "sdO" is another spectral type not contemplated in the original MK system. Nonetheless, sdO is a recognized spectral type and denotes a hot subdwarf (hence "sd"). Types **TT** = 30-33 (containing "sd") and **TT** = 25 (containing "CHd") are occasions for which lower case is used to denote part of a spectral (temperature) type string. Generally, lower case letters are used to modify luminosity classes, as in "Va" and to describe peculiarities.

- L5

    L5 → 17.15.00.0000. Our encoding system recognizes the need for newly defined cool spectral types, such as the transition from stars to brown dwarfs, as based on infrared spectra. Hence, we provide for types L and T.

- SC9/8 Tc

    SC9/8 Tc → 26.19.00.0041. The S spectral type is a supplementary type to the MK system. It was introduced to describe spectra of cool stars with nonsolar abundance peculiarities (e.g., [C] > [O] and enrichments of nucleosynthetic s-process elements like Zr. The digit "9" is the spectral subtype, and it maps to value 19. Note in this example that the integer 8 (on a scale of 1 to 10) parameterizes the elemental abundance ratio C/O in the S-type stars. (This fact is not part of this spectral class encoding system!) The SC and Tc anomalies are handled by $P_3P_4$ = 41 peculiarity designations to the spectral type S (Table 3).

- kA3hF0mF2 III

    kA3hF0mF2 III → F0m III → 13.10.23.0005. This is an example of a detailed spectral classification for an Fm star, where the designation "m" for "metallic lined" requires that $P_4$ = 5 for an F-type star (Table 3). Spectral classifications of these stars are given three distinct subtypes, one each for the K-line ("k"), the hydrogen lines ("h"), and metallic lines ("m"). As noted above, such notations can precede the associated subtype. Only the

hydrogen line type gives the true spectral type, which in this case is F0 (**TT** = 13, **tt** = 10). See Appendix A for a fuller description.

Bep → 11.00.00.3200.  Bep denotes a Be star. By definition the spectrum shows, or has shown in the past, Balmer emission lines. The "p" indicates the spectrum has peculiarities of an unspecified nature. In this case **P$_1$** = 3 and **P$_2$** = 2 because the peculiarity order is maintained when both p and e are specified. The precedence of e over p (or vice versa) is the only exception to the rule for **P$_1$** and **P$_2$** that their placement in the map string is determined by the order of the peculiarity in the **P$_1$P$_2$** table.

- Bpe
Bpe → 11.00.00.2300. Bpe denotes a Bp star. Their spectra typically are peculiar in the strengths of the He I absorption lines. In this case, the Bp spectrum also happens to show Balmer emission. (In contrast, Be stars show normal He I line strengths.) The precedence of the peculiarities e and p occurring together, is determined by the first-occurring symbol in the classification string, so the numerical **P$_1$P$_2$**-code is reversed from the previous example: **P$_1$** = 2 and **P$_2$** = 3.

### 5.2 Composite spectra

Composite spectra contain lines of both binary components, and therefore some estimate can be made of both their spectral types. However, the identification of two stars in a single composite spectrum requires the use of at least 3 of the **PPPP** codes. The code **P$_1$** = 1 designates a composite spectrum. If one additional global peculiarity is noted in the primary star's spectrum, it can be encoded in **P$_2$**. Any peculiarity noted for the secondary is ignored, and any type-dependent peculiarity for the primary is ignored.  If the spectral type can be discerned **P$_3$** and **P$_4$** denote the spectral type of the secondary star; no subtype, class, and/or peculiarity is coded  for the secondary. If the secondary spectrum is unknown **P$_3$** and **P$_4$** are set to zero.

- B5e+sd0
B5e+sd0 → 11.15.00.1331. Like most examples of composite spectra, no luminosity information is available in this case. **P$_1$** = 1 announces the composite nature of the  spectrum. The peculiarity code **P$_2$** is set to 3 to show that the spectrum of the primary star exhibits emission lines (by definition, the component that contributes the most optical flux). The spectral type of the secondary determines the **P$_3$ P$_4$** codes, which are now used for its spectral type, sd0.

- WN5ha+OB
WN5ha+OB → 52.15.00.1056 WN5 maps to TT=52 and tt = 15. In this case because the spectral type of the secondary is given, **P$_3$P$_4$** becomes the TT value for an OB star (i.e., 56).  The "ha" string is discarded because the object is a composite system.

- sdB3e:+M4e

sdB3e:+M4e → 32.13.00.1316. Here the primary's spectral type and subtype are sdB3, hence **TT.tt =** 32.13. As in previous examples, the spectrum is composite. Because "composite" occurs first in the $P_1/P_2$ table, it is $P_1$, and it is coded to the "composite" value of 1. The character ":" is ignored. Because lines of the primary are in emission, we have $P_2$ = 3. The known spectral type of the secondary, M, determines the (merged) $P_3P_4$ value of 16.

- Gp + M0III

Gp + M0III → 14.00.00.1216. This case demonstrates the ability of our coding system to allow for the two most important global peculiarities, which are, respectively, "composite" and "p." Note, incidentally, that for the complementary examples, "Gpe + M0III" and "Gep + M0III", $P_1P_2$ would equal 12 and 13, respectively. Notice also that the LL code is reserved for the luminosity class of the primary, which is unspecified in these examples. Thus, **LL** =00.

Finally, we note that the sole spectral designation "composite" or "comp." is to be discarded, as it adds information to a non-existent spectral type TT.

## **References**

[1] M. A. Smith, R. Gray, C. Corbally, I. Kamp, R. Thompson, 2007, *A Spectral Class Encoding System: Mining Data and Metadata in the 21st Century*, AAS, 211, 4709S, http://adsabs.harvard.edu/abs/2007AAS…211.4709S

[2] R. O. Gray, C. Corbally, Stellar Spectral Classification, Princeton Series in Astrophysics, 2009

[3] Skiff, B. 2007, yCat….102023S [consult VizieR catalog]

# Appendix A: Notes on treatment of special Am, Fm, and similar classifications

In this standard we use regular expressions to define 6 groups of "special" spectral classifications for Am stars, Fm stars, and other (possibly unrelated) stars with similar syntax. Stars described with the usual spectral classification syntax (e.g., A5m) are parsed like any other spectral type.

The following notation is used:
X, X', X" = spectral types (i.e., single upper case letters such as O,B,A…)
s = spectral subtype (i.e., a single digit or a combination of digits such as 7.5 or 6/7). Note that multiple occurrences of subtypes are independent and can therefore have different values.
L = luminosity class (i.e., a combination of I,V,a,b and/or z characters). Note, as for spectral subtypes, that multiple occurrences are independent.
( ) = parentheses are used to indicate the spectral classification used in the final encoding. Additional peculiarities found in the expression will be interpreted according to this spectral type. (Note that some groups automatically assign certain peculiarities.)

All other characters shown below represent themselves. Any characters (e.g., peculiarities) can follow the listed strings, but none may precede them. In general, the "s" and "L" characters are optional.

The six groups of "special" spectral classifications with defined peculiarities are:

**Group 1:** All 3 standard abundances (h, k and m) listed or implied.
   Four formats are allowed:
   Xs/(X'sL)/X"s,  kXsLh(X'sL)mX"s,  knXsLh(X'sL)mX"s,  h(Xsl)kX'smX"s
   $P_4$ = 5  (Am or Fm)

**Group 2:** Abundances h and m, and g abundance cases: hg, hgm, and hgkm
   Four formats are allowed:
   h(XsL)mX's,  h(XsL)gX'smX"s,  h(XsL)gX'skX"smX"'s,
   h(XsL)gX'skX"smX"'s
   $P_4$ = 5 (Am anomaly), except:
   If the "m" spectral type is earlier than the h or g spectral type, then $P_4$=3 (metal weak).

**Group 3:**  Abundances k and h.
   Two formats are allowed:
   kXsh(X's),  knXsh(X's),
   if k TT value > h TT value, $P_4$=4 (metal rich)
   if k TT value > h TT value, $P_4$=3 (metal weak)
   if above are equal, and the k tt value >  h tt value, then  $P_4$=4 else $P_4$=3

**Group 4:** Abundances k and m, and m with two spectral types.
  Four formats are allowed:
 k(XsL)mX's, kn(XsL)mX's, k(XsL)m, (XsL)mX's
 $P_4 = 5$.

**Group 5:** Single abundance h or k designations, and cases where A spectral types are
  followed by a k abundance. Three allowed formats:
 h(XsL), k(XsL), (AsL)kXs
  No default value is assigned for $P_4$.

**Group 6:** He abundance cases including khHe, hHe, and kHe.
   Four formats are allowed:
  kXsLh(X'sL)HeX''s, h(XsL)HeX's, k(XsL)HeX's, knXsLh(X'sL)HeX''s
  if the TT value for the extracted spectral type is ≤11 (i.e., O and B stars), then $P_3=1$.

# Appendix B: Examples of Implementation

Although the core of this Note refers to the encoding system *per se*, it is helpful to give examples of how the mapping scheme might be implemented by data providers, both in terms of the routing of requests and the presentation of a web interface form for users. This appendix is not part of the classification standard defined in the main text body.

## *B1. Data Source*

Data providers will want to utilize a resource that provides a "best" spectral type for all objects for which pointed spectroscopic observations exist in its archives.

One such service is the Sesame service made available at CDS/Simbad at:

http://webviz.u-strasbg.fr/cgi-bin/Sesame

The following url would retrieve CDS's "best" spectral type, B2III, for HD60753:

http://vizier.u-strasbg.fr/viz-bin/nph-sesame/-ox/?HD60753 .

Another data source is the current Skiff catalog, which is available from VizieR.

at http://vizier.cfa.harvard.edu/viz-bin/VizieR?-source=B/mk .

In MAST's (Multimission Archive at Space Telescope Science Institute) implementation a script may harvest and update periodically new spectral types for all stars in its table of representative spectra. A parsing program, encapsulating the scheme and encoding rules, discussed in Sections 3, 4, and the appendices, is then run on these spectral types. The encoded classes, in **TT.tt.LL.PPPP** format, are stored as integers in seven fields in the database – one each for **TT**, **tt**, **LL**, and the four **P** codes.

## *B2. Implementation of Database search queries*

Although most of this document is devoted to the exposition of the spectral class nomenclature system itself, i.e. the mapping from spectral classification strings to a numerical coding system, important details must be resolved to ensure the success of the system to archival data users. Such details may be resolved differently by different archive centers, depending on the convenience and practices of members of their particular user communities.

For example, one need that we foresee is the ability to easily request a "bundle" of peculiarity types that are related astrophysically. Thus, one might want to download lists of stars and their data that are peculiar in certain similar respects, e.g. "give me a list of A-type stars that are Ap, whether the 'p' occurs in the spectral classification string or not." This query would select all stars classified as "A SrCrEu stars" (with permutations among these anomalies) and "Si stars" that are by astrophysical definition also Ap stars, in addition to those with explicit classification of "Ap." A similar query might include mutually exclusive parameters that describe extreme stellar conditions, such as "give me all B stars that are either sharp-lined ("s") or broad-lined ("n").   Decisions as to how or whether to permit such questions permit the nomenclature system to retain its versatility without overburdening it with rules for the peculiarity codes that can be difficult to interpret and are susceptible to errors in searches and string parsing.

A second class of queries might utilize the IVOA Simple Spectral Access Protocol (SSAP) to retrieve information on stars of a particular spectral type from data centers supporting the new spectral class system. For example, by defining a new SSAP input parameter SPCLASS, users could search for all "G2e" stars by adding the string "SPCLASS=14.12.3000" to their SSAP request. Some range-list type queries should also be possible. Then,  a request to search for both A3II and A3IV stars with no known peculiarities could be parsed as "SPCLASS=12.13.19.0000;12.13.27.0000".  Ranges could be allowed if performed on one element at a time.

Although much work needs to be done in these areas, the authors have made several provisional decisions for its own web-based implementation of the spectral class system at MAST. For example, we may allow multiple selections on the peculiar fields.

### *B3. Web Interfaces*

The following lists examples of interface forms for spectral classes that MAST might implement. Search results can include a star name (e.g. "HD12345"), relevant metadata, the **TT.tt.LL.PPPP** code, the original spectral classification string, and a link to download one or more representative datasets as a second step in the data discovery process.

#### B3.1 Interface 1:  A standard query form

Interface 1 allows users to specify as little or as much of the spectral object class ranges as they wish. When the user selects a particular spectral type the appropriate 4-table set of peculiarity menus appear. If the user wants to retrieve all A stars, for example, he/she clicks only on type "A" and submits the query.

If a user wants only "normal " A stars the four peculiarity values will be specified as "none" in our example. We show as an example the **$P_1P_2P_3P_4$** menus that appear as defined peculiarities for spectral type A. One may highlight any set of selections on the **tt**, **LL**, and **PPPP** menus as needed (except across spectral types if spectral subtypes are also specified. Submitting a request for B1-B9 is allowed, but a selection of B9-A0 requires requests, first for B9, and then A0.

*Figure 1: A MAST web search form allowing users to input a range of spectral classes, including spectral types, luminosity classes and any peculiarity codes. In this case the A spectral type has been highlighted by mouse action. This event causes the $P_3P_4$ menus to be highlighted with peculiarities appropriate to type A. This page can be found at http://archive.stsci.edu/spec_class/search.php .*

### B3.2 Interface 2: user defines request through type/class plane

Figure 2 is a mock-up of a color-coded "HR Diagram" contour plot that could allow users to define a region of **TT.tt.LL** space by a quick cursor drag/click action. This form allows one to request in one action all objects within a range of spectral types and luminosity classes, irrespective of peculiarity conditions. The box specified may cross TT boundaries (e.g., B9-A0). Our mock up attempts to exhibit population demographics of the stars in a database for pointed spectroscopic missions. Some classes, like Ap stars, can appear as groups. This would make it easier to select all the members of such classes.

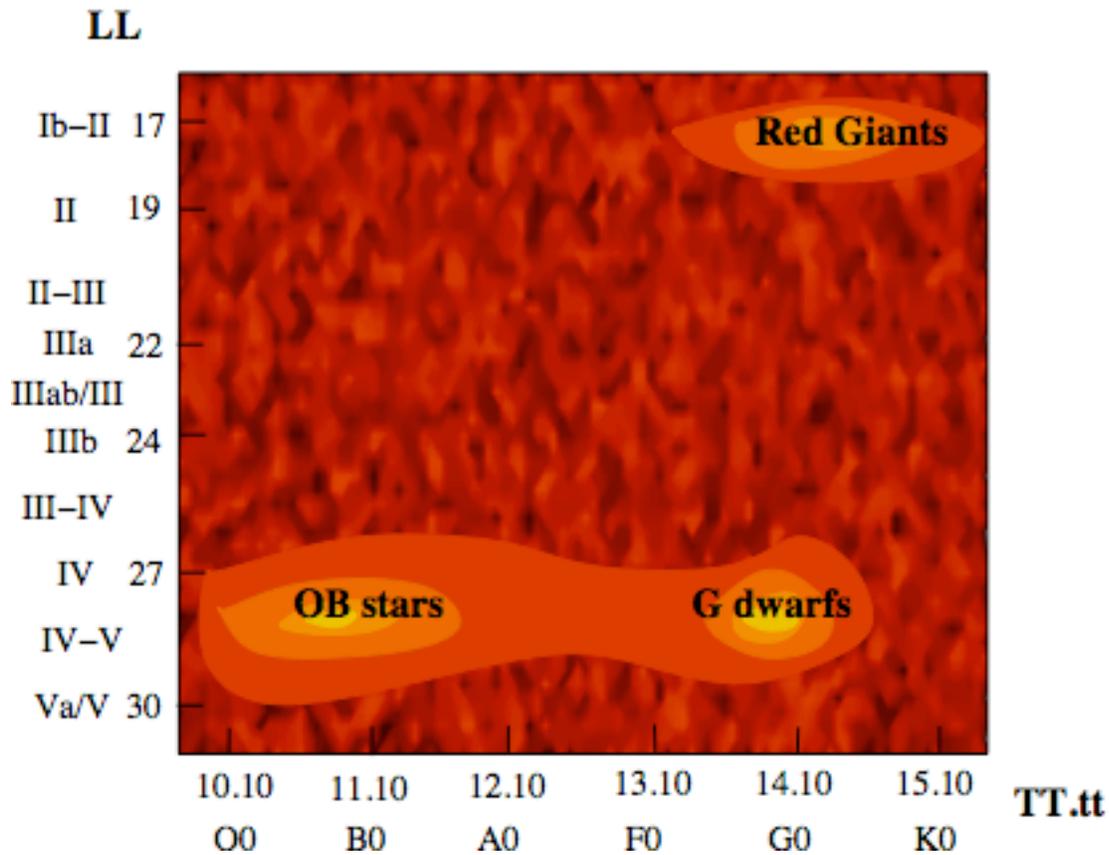

*Figure 2: A user may request a list of stars over a range of spectral classes for retrieval by drawing a cursor box over a dynamic TT.LL plane on a web portal. This mode will retrieve all stars irrespective of peculiarity status.*

# Appendix C: A supplementary "detail" flag

As a matter of implementation, the authors have found that a "detail" flag ("d") can be useful for defining various levels of over- or under-specification of a star's full spectral type. The flag values have the following definitions:

- **d = 0 :** This means that a classification has been successfully interpreted in our scheme, but some characters may have been ignored, as noted in the standard.

- **d = 1 :** The original spectral classification of the star included a decimal spectral type, e.g. B0.2. The decimal subtype was truncated in parsing; for decimal subtypes 9.5-9.10 of hot stars (see **tt** section of Table 1) the truncation is to subtype 9.5. The user may recover the original (untruncated) decimal subtype in various ways (e.g. through a display of the subtype of the original classification on the results page).

- **d = 2 :** A flag set to this value indicates that the classification in the literature was encoded, but was considered "slightly ambiguous". This serves as a warning to the user that the classification might need to be looked at more carefully. Thus far, the only example of this case is when the spectral classification of an A or F star is given solely from the Ca II K line, e.g., "kA3." In such a case, this spectral type was interpreted as the hydrogen line type A3.

- **d = 3 :** The spectral type could not be interpreted using the described nomenclature system. Classifications in this category would include: "symbiotic", "em", "PN", "composite", etc.

The d flag is not part of the formal standard defined in the body of this document.

As a practical matter, we note that more than one classification may exist for a given star in the literature. The archiving agency should decide the precedence of how to choose among different spectral types given for objects in its database. (MAST has made a decision based on dates of publications and a hierarchy of catalogs. Information on this subject will be provided upon request.)